# Using Design Sketch to Teach Bubble Sort in High School*

Chih-Hao Liu[1], Yi-Wen Jiu[2], and Jason Jen-Yen Chen[3]
[2]Wu-Ling Senior High School
Tao-Yuan, 33059 Taiwan
[1, 3]Department of Computer Science and Information Engineering
National Central University
Jhong-Li, 32001 Taiwan

**Abstract**— Bubble Sort is simple. Yet, it seems a bit difficult for high school students. This paper presents a pedagogical methodology: Using Design Sketch to visualize the concepts in Bubble Sort, and to evaluate how this approach assists students to understand the pseudo code of Bubble Sort. An experiment is conducted in Wu-Ling Senior High School with 250 students taking part. The statistical analysis of experimental results shows that, for relatively high abstraction concepts, such as iteration number, Design Sketch helps significantly. However, it is not so for low abstraction concepts such as compare, swap, and iteration.

## 1 INTRODUCTION

PROGRAMMING is a new subject for high school students in Taiwan. Unlike the demand of memorization of the liberal arts and the logical reasoning of mathematics and physics, programming is a nimble and flexible activity that transforms concrete objects into abstract concepts. It does not demand one single answer to the problem posed to students. Answers depend upon how students approach the problem. Programming features no fixed solution, and that is the reason why it is difficult to teach.

Sorting is the most common material in program design. It has several methods, among which Bubble Sort seems the easiest. Still, teachers have been encountering difficulties having all students understand its principle. Usually, high school teachers lecture on Bubble Sort by means of array which includes two concepts, position (index) and content (value). Swaps are made resulting from comparing contents at some specified positions. When comparisons and swaps are repeatedly made, students will no longer remember what the content of a position is. This is because that some concepts are easier to comprehend by human brain, but some are not. This paper thus studies which concepts facilitate teaching Bubble Sort.

We used PowerPoint demonstration of Design Sketch [4] of Bubble Sort in our teaching. The demonstration aims to ease the concept comprehension of Bubble Sort.

## 2 LITERATURE REVIEW

Based on Cognitive Informatics [8], there are five levels of information processing in human brain from the bottom level of analog objects to the top level of philosophies, in which programming language is located between the third level and the fourth (figure 1). We figure that pseudo code is a bit higher than programming language in the hierarchy

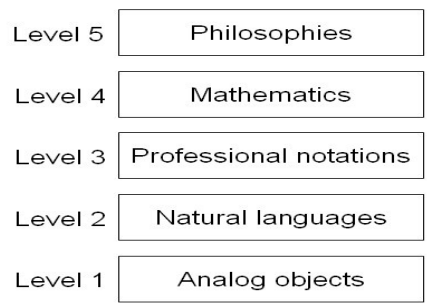

Fig 1. Levels of human information processing

Cognitive Informatics also shows that human perception can be divided into two worlds: physical (or concrete) world and abstract (or perceived) world. Humans perceive sensor data in the physical world through their senses and then conceptualize them into the abstract world (figure 2). Programming can be likewise considered as a process of this kind.

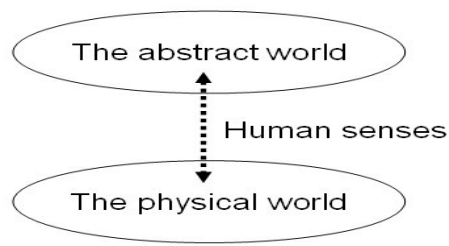

Fig 2. Abstract world and Physical world

---

Piaget's Schema theory tells us that each individual understands the world around him in accordance with his basic, inner behavior pattern. When new concept that this individual acquires encounters the formerly acculturated one, assimilation and accommodation will take place to be adjusted to the new environment [3]. The amount of time needed for this depends upon the level of conceptualization of each individual. However, some people will never comprehend certain concepts; that is to say, for some people, the conceptualization mentioned above may not likely occur, while, for others, it does [6]. The methodology of Programming can thus be different from one person to another.

## 3 EXPERIMENTAL DESIGEN

The following is divided into three sections. Section I details the regrouping of the students involved in this experiment. Section II delineates the basic training required of the students. Section III explains the Bubble Sort Design Sketch and the experimental procedures.

### 3.1 Regrouping Students

In order to be more representative with the population for this experiment, six classes of all first-year students were selected at random of Wu-Ling Senior High School, and then regrouped into experimental group and control group. There are three classes in each group. The experimental group consists of 123 students. And the control group 127 students. There are 250 students all together. Before the experiment is conducted, those students are required to take the Computer Self-Efficacy Scale [5] to make sure these two groups of students are at the same proficiency level. The statistical T Test shows the positive result that the two groups share the same score.

### 3.2 Basic Training of Programming

Before participating in the experiment, most of the students are inexperienced with programming. This explains the necessity of involving both groups of students in a basic training for one hour a week, six weeks consecutively. The emphasis of the training is on the basics of the Object Oriented Programming (Object, Attribute, Method, Event), Data Type, Variables and Constants, Program Structures (Sequence, Select, Iteration), Array and Function.

One identical lecturer will be responsible for the training in order to maintain the impartiality of the content and the way of the lecturing for both groups, so that Pygmalion Effect[φ] can be avoided [2]. Besides, we deliberately keep the students ignorant of their regrouping (i.e. each student does not know which group he or she is assigned to), thus not to reinforce John Henry Effect[θ] [1].

### 3.3 Bubble Sort Sketch and Experimental

---

[φ] This term is originated from the Greek Mythology. It means that any lecturer can be partial for the favor of the experimental group for the experiment result.
[θ] This term refers to the fact that during Industrial Revolution, an American worker named John Henry competed with machines out for the fear of being laid off. In this experiment, the control group will do their outmost to compete with the experimental group.

### Lecturing

This section is sub-divided into two parts: the first describes the concepts of the Bubble Sort Design Sketch and the second details the procedure of the experiment.

#### 3.3.1 Design Sketch

We do Design Sketch using PowerPoint for the experimental group. The sketch contains six concepts: value, position, compare, swap, iteration, and iteration number.

1. Value: The size of a bubble is marked by a number called value. The bigger the size, the larger the number is, as illustrated in figure 3.

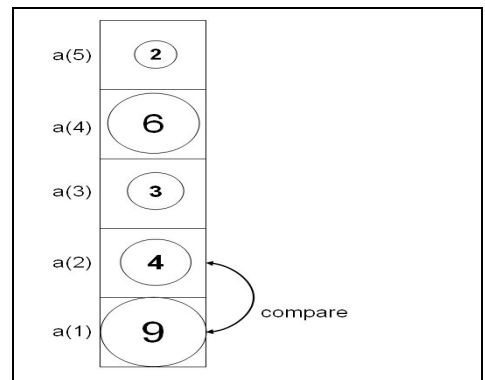
Fig 3. Concepts of value, position and compare

2. Position: Bubbles are vertically arrayed in the boxes. The position of the boxes from top to bottom indicates the ordering of the amount of value from large to small, as illustrated in figure 3.

3. Compare: A two-arrowed curve points to two different numbers, which are marked by two different colors, as illustrated in figure 3.

4. Swap: When the requirement is met for comparison (that is, the bubble underneath is bigger than the one above), we animate the process of swap, as illustrated in figure 4.

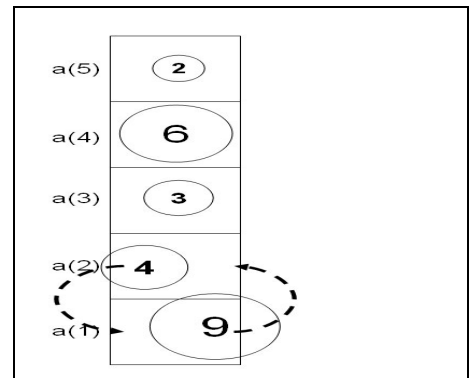
Fig 4. The concept of swap

5. Iteration: When the biggest bubble floats to the top by means of compare and swap, it means one round of sorting is completed and we mark "completed" on the top of the box. This shows the concept of it-eration, as illustrated in figure 5.

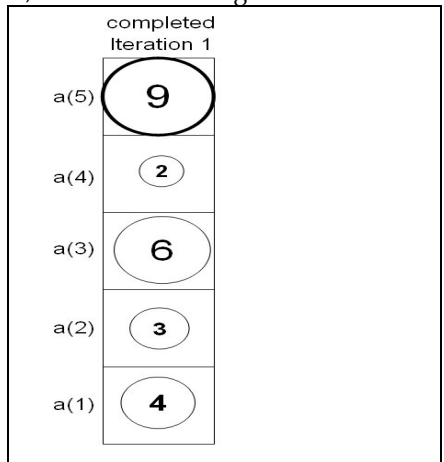

Fig 5. The concept of iteration

6. Iteration Number: Each iteration resulted from sorting is shown in the Design Sketch, indicating the n-1 completion of iteration, as illustrated in figure 6.

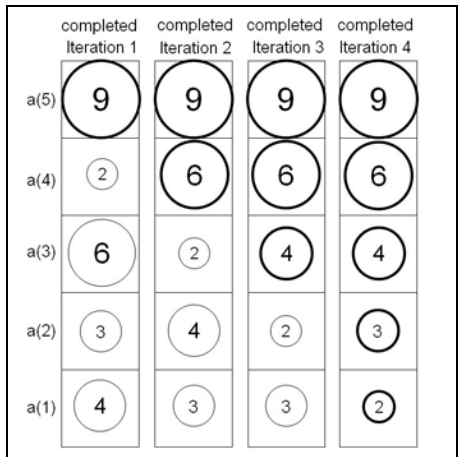

Fig 6. The concept of iteration number

There are different levels of human conceptualization. The comprehension of the lower level of abstraction is the basis for that of the higher one [7]. For example, concrete concepts like "bird" and "fish" belong to the lower level of conceptualization and are easily comprehensible, whereas the concept of "animal" is relatively abstract than the former two. Bird and fish are thus the necessary basics for human to build up the comprehensive concept of "animal." And "animal" can be likewise the prerequisite concept for further understanding of a highly abstract one "being," as illustrated in figure 7.

The six concepts of Bubble Sort represent different levels of conceptualization. Concepts like "value," "positioning," and "compare" are concrete and more comprehensible, all of which serve as the necessary groundwork for further abstract concepts like "swap" and "iteration." The ultimate level of abstraction "iteration number" necessitates the well-conceptualized first five, as illustrated in figure 8. A precise description of the concepts should be done using an ontology language.

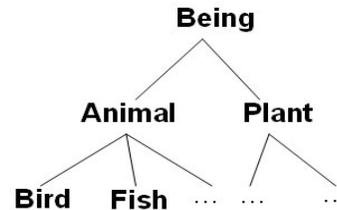

Fig 7. The example of conceptualization levels

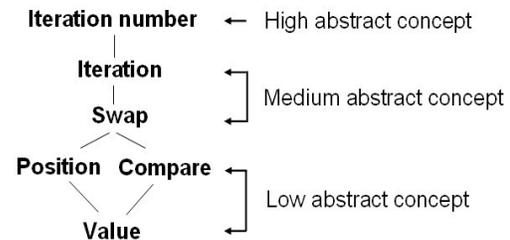

Fig 8. The conceptualization levels of Bubble Sort

### 3.3.2 Experiment

The experiment can be conducted when the basic training is completed. For the students in the experimental group, they are first lectured on with the animated Design Sketch. The sketch aims at animating the well-understood phenomenon of bubble floating to water surface to serve as a "metaphor" for students to precisely understand the program logic. As a matter of fact, Bubble Sort is named after this great metaphor of floating bubble. Then the position of the bubble corresponds to the value of the number, which is well put into the array data structure. Lastly, the six concepts of value, position, compare, swap, iteration, iteration number are mapped into the program constructs of selection structure and iteration structure, and then are transformed into pseudo code, as illustrated in figures 9 and 10. On the other hand, the students in the control group are likewise lectured on with the same procedures, but without the animated Design Sketch.

The lecturer observed that the students in the experimental group showed more interest about the lecture. It seems that animated Design Sketch attracts attention of the students with much success. Consequently, they are more willing to think for themselves and volunteer to answer the guided questions posed by the lecturer. Besides, the brainstorming involves more their neighboring counterparts into their discussion. On the other hand, students in the control group are relatively inactive because the lecture they received is much more orally based without animated simulation.

### 3.4 Evaluation

The experiment is followed by an evaluation. The evaluation (Table 1) is a questionnaire, in which four questions are listed, 2 points each.

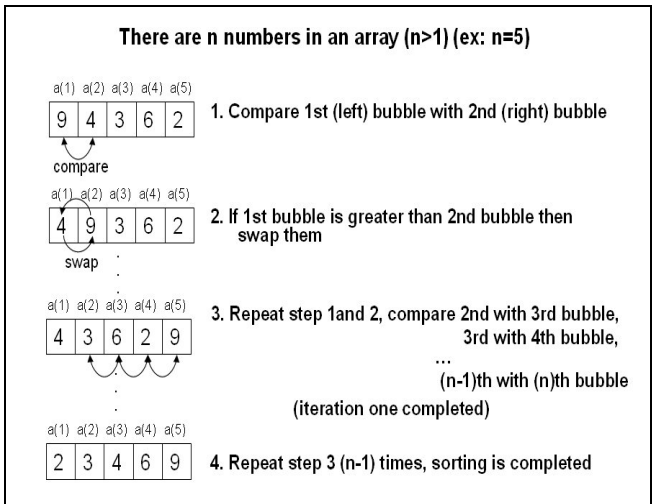

Fig 9. Bubble Sorting with array structure

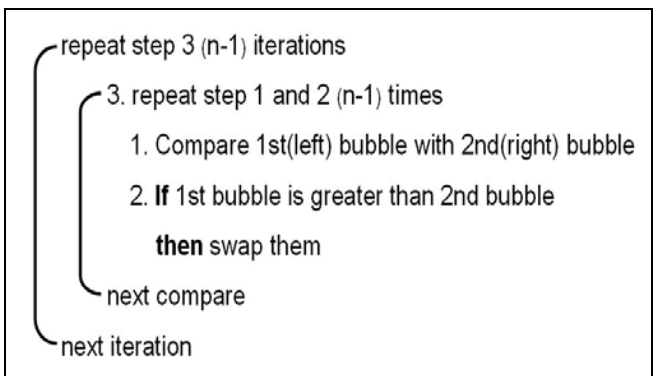

Fig 10. Pseudo code

The reason for this evaluation on the six basic concepts—value, position, compare, swap, iteration and iteration number—is to determine if the students really understand the logic of the pseudo code.

Table 1 the Evaluation

| Item number | Questions |
|---|---|
| 1 | A series of numbers: 21,19,37,5,2. Arrange the numbers from small to large, using Bubble Sort. When the first swap is completed, the order of these numbers is ______________. |
| 2 | A series of numbers: 4,2,6,5,1,7,3,8. Arrange the numbers from small to large, using Bubble Sort. When the first iteration is completed, the order of these numbers is ______________. |
| 3 | How many times of iteration is needed to complete an array of 8 elements when using Bubble Sort? ______ |
| 4 | A series of numbers: 4,2,4,3,1. Arrange the numbers from small to large, using Bubble Sort. Which two numbers are involved in the second compare of the first iteration? ______ |

We thus hypnotize that, through the experiment, the students of the experimental group are more capable of comprehending these six concepts than those in the control group.

## 4 EXPERIMENTAL RESULTS

The result of the evaluation is analyzed with the statistical T Test to verify our hypothesis. We compare the average score of each identical question by the two groups. Table 2 shows the result.

Table 2 the Result of Evaluation

| Item Number | concept | Average score of the experimental group | Average score of the control group | P value |
|---|---|---|---|---|
| 1 | Position Compare Swap | 0.881 | 1.296 | 0.0005*** |
| 2 | Position Compare Swap Iteration | 1.034 | 1.424 | 0.0008*** |
| 3 | Position Compare Swap Iteration Iteration number | 1.881 | 1.664 | 0.0039*** |
| 4 | Position Compare Swap Iteration | 1.220 | 1.536 | 0.0038*** |

*** $p < 0.01$

The experimental group scores better than the control group in Question 3. This question is about the concept of iteration number and with the analysis of the T Test, the P value comes to 0.0039 (<α(0.01)), which clearly indicates that the students' performance reaching a significant level, as illustrated in table 3. The result shows that the Design Sketch is very helpful in enhancing students' comprehension of a highly abstract concept, namely, iteration number.

The experimental group scores poorly than the control group in Question 1, 2, and 4. Through the T Test, the results reach the significant level, which means that Design Sketch obstructs the understanding of low abstract concept.

## 5 CONCLUSION

We previously assumed that the animated Design Sketch helps produce better comprehension of Bubble Sort. The experimental group is supposed to score better than the control group. However, what we have obtained from the experiment does not meet our expectation. Here is our analysis:

Table 3 the Students' Performance Reaching a Significant Level

| t-Test: Two-Sample Assuming Equal Variances | | |
|---|---|---|
| Question 3 | Experimental group | Control group |
| Mean | 1.8814 | 1.664 |
| Variance | 0.2251 | 0.5636 |
| Observations | 118 | 125 |
| Hypothesized Mean Difference | 0 | |
| Df | 241 | |
| t stat | 2.6799 | |
| P(T<=t) one-tail | 0.0039 | |
| t Critical one-tail | 1.6512 | |
| $H_0$: the score of the experimental group equals to the score of the control group $H_1$: the score of the experimental group is greater than the score of the control group Conclusion: do not reject $H_1$ | | |

The emphasis of Question 3 is on iteration number, where the experimental group scores better than the control group. Thus we have inferred that the animated Design Sketch clearly explains the concept of "iteration number," whereas the array method is not able to yield the same result. That is the reason why the experimental group performs better than the control group.

Question 1, 2, and 4 put more emphasis on medium and lower levels of abstraction, such as "compare," "swap," "position," and "iteration." Although the animated Design Sketch is more advantageous of elucidating concepts like "compare" and "swap," these two concepts is not highly abstract and they can be similarly made clear to the control group simply by orally based lecturing. When it comes to the concept of "position," the students of the experimental group are lectured on with the animated Design Sketch, in which the bubbles arise floating vertically from bottom to top. The students have to convert the vertical reasoning to the horizontal array of position from left to right. Most of the students fail to assimilate these two modes of reasoning with much success. The students consequently are not able to fully understand the Bubble Sort. On the other hand, the students in the control group are free from the obstruction of the animation. They are lectured on with the array: that is, the ordering of numbers is nothing but the same ordering of the array. They encounter no vertical-horizontal assimilation. That's why the control group scores better than the experimental group in Question 3.

We thus conclude that, without the animated Design Sketch, human brain is capable of understanding low abstract concepts such as "value," "position," "compare," "swap," and "iteration" in the Bubble Sort. However, when it comes to high abstract concepts such as "iteration number," the animated Design Sketch is proven very helpful.

When we designed this experiment, we added the horizontal array in the lecture on the experimental group. This may have resulted in unnecessary assimilation which obstructs the students' comprehension of the vertical bubble floating. Further, "left" and "right" in the pseudo code in figure 10 reflect horizontal array. If there be any experiment of this kind in the future, the array should be completely eliminated. This may help comprehend all the concepts in the pseudo code. Besides, this research employs low-level array data structure. In the future, more complicated, high-level data structures such as tree and graph may be included to see if they are helpful.

## ACKNOWLEDGMENT

The authors would like to thank the National Science Council of Taiwan for the supports.

**Chih-Hao Liu** received his Master degree of Information Engineering from Chaoyang University of Technology. He is currently a PhD candidate in the National Central University in Taiwan. He joined the software engineering laboratory in 2005. He also participated the SIM (Service-oriented Information Marketplace) project from 2005 to 2007. And, his current research interests focus on Semantic Web and Agent.

**Yi-Wen Jiu** has been teaching in Wu-Ling Senior High School, Tao-Yuan, Taiwan since 1990. He received his M.S. degree in computer science and information engineering from National Central University, Jhong-Li, Taiwan, in 2007. His research interest is cognitive informatics.

**Jason Jen-Yen Chen** is with the Department of Computer Science and Information Engineering in National Central University in Taiwan. He earned international recognition by winning Top, Third, and Fifth Scholar in the world in the field of System and Software Engineering in 1995, 1996, and 1997, respectively. The ranking is based on cumulative publication of six leading journals in that field. His current research interests include agile method and agent technology.